\documentclass[a4paper,11pt]{article}

\usepackage{jinstpub} 

\title{\boldmath Tracker-In-Calorimeter (TIC): a calorimetric approach to tracking gamma rays in space experiments}

\author[a,b]{O. Adriani}
\author[c]{G. Ambrosi}
\author[d]{P. Azzarello}
\author[e]{A. Basti}
\author[a,b]{E. Berti}
\author[c,f]{B. Bertucci}
\author[e,g]{G. Bigongiari}
\author[b]{L. Bonechi}
\author[a,b]{M. Bongi}
\author[b]{S. Bottai}
\author[b]{M. Brianzi}
\author[e,g]{P. Brogi}
\author[b,h]{G. Castellini}
\author[c,f]{E. Catanzani}
\author[e,g]{C. Checchia}
\author[a,b]{R. D'Alessandro}
\author[b]{S. Detti}
\author[c]{M. Duranti}
\author[b,i]{N. Finetti}
\author[c,*]{V. Formato}
\author[c]{M. Ionica}
\author[e,g]{P. Maestro}
\author[b]{F. Maletta}
\author[e,g]{P. S. Marrocchesi}
\author[b,1]{N. Mori\note{Corresponding author.}}
\author[b]{L. Pacini}
\author[b]{P. Papini}
\author[b,h]{S. Ricciarini}
\author[c,f]{G. Silvestre}
\author[b]{P. Spillantini}
\author[b]{O. Starodubtsev}
\author[e,g]{F. Stolzi}
\author[e,g]{J. E. Suh}
\author[e,g]{A. Sulaj}
\author[b]{A. Tiberio}
\author[b]{E. Vannuccini}

\affiliation[a]{Department of Physics and Astronomy, University of Florence, via G. Sansone 1, Sesto Fiorentino I-50019, Florence, Italy}
\affiliation[b]{INFN sezione di Firenze, via G. Sansone 1, Sesto Fiorentino I-50019, Florence, Italy}
\affiliation[c]{INFN sezione di Perugia, via A. Pascoli, Perugia I-06123, Italy}
\affiliation[d]{D\'epartement de Physique Nucl\'eaire et Corpusculaire, University of Geneva, 24 quai Ernest-Ansermet, Geneva CH-1211, 
Switzerland}
\affiliation[e]{INFN sezione di Pisa, largo B. Pontecorvo 3, Pisa I-56127, Italy}
\affiliation[f]{Department of Physics and Geology, University of Perugia, via A. Pascoli, Perugia I-06123, Italy}
\affiliation[g]{Dipartimento di Scienze Fisiche, della Terra e dell’Ambiente, University of Siena, via Roma 56, Siena I-53100, Italy}
\affiliation[h]{IFAC-CNR, Via Madonna del Piano 10, I-50019 Sesto Fiorentino (Firenze), Italy}
\affiliation[i]{Department of Physical and Chemical Sciences, University of L’Aquila, Via Vetoio, I-67100 L’Aquila, Italy}
\affiliation[*]{now at INFN sezione di Roma Tor Vergata, Rome I-00133, Italy}
\emailAdd{mori@fi.infn.it}

\abstract{A multi-messenger, space-based cosmic ray detector for gamma rays and charged particles poses several design 
challenges due to the different instrumental requirements for the two kind of particles. Gamma-ray detection requires layers of high Z 
materials for photon conversion and a tracking device with a long lever arm to achieve the necessary angular resolution to separate point 
sources; on the contrary, charge measurements for atomic nuclei requires a thin detector in order to avoid unwanted fragmentation, and a 
shallow instrument so to maximize the geometric factor. In this paper, a novel tracking approach for gamma rays which tries to reconcile 
these two conflicting requirements is presented. The proposal is based on the Tracker-In-Calorimeter (TIC) design that relies on a 
highly-segmented calorimeter to track the incident gamma ray by sampling the lateral development of the electromagnetic shower at different 
depths. The effectiveness of this approach has been studied with Monte Carlo simulations and has been validated with test beam data of a 
detector prototype.}

\keywords{Gamma telescopes, Particle tracking detectors}

\begin{document}
\maketitle
\flushbottom

\section{Introduction}
\label{sect:introduction}
A multi-messenger detector for cosmic rays, capable of detecting charged particles and gamma-ray photons, represents an appealing 
perspective for the next generation of space-based, high-energy direct detection experiments from both the scientific and the logistic 
points of view. Such a facility would provide invaluable data for various research fields, at a significantly reduced cost compared to a 
classic two dedicated detector solution (one for gamma rays and one for charged particles). Its design however poses many challenges. 
In order to make precise measurements with a good statistical significance also at high energies, scientific requirements range from good 
energy and tracking resolution to high geometric factor and effective area. Trying to meet all these requirements for charged particles and 
gamma rays with a single detector raises some inevitable conflicts.

Gamma-ray tracking is typically performed using a pair-conversion telescope. Each plane of the telescope is made of a layer of passive 
material (e.g. tungsten) in which the primary photon is converted into an e$^+$-e$^-$ pair, coupled with a sensitive layer (e.g. silicon 
microstrip detector) which measures the impact position of the electron and the positron. Several of these planes are stacked to increase 
the photon conversion probability and to provide multiple position measurements along the e$^+$-e$^-$ trajectory; from these measurements 
the track of the incident photon is reconstructed. The planes are suitably spaced apart from each other, so as to achieve the lever arm 
that, in combination with the spatial resolution of the position measurements, provides the required tracking resolution. This concept has 
been successfully implemented in recent space-borne gamma-ray detectors, like Fermi \cite{Atwood2007} and AGILE \cite{Bulgarelli2010}, and 
can be considered as the reference technique for gamma-ray tracking at energies above some hundreds MeV. Being sensitive to ionization, such 
a device is in principle also able to track charged primary particles and is actually used in detectors for charged particles with gamma-ray 
detection capabilities like DAMPE \cite{Chang2017}

On the other hand, charged particle physics requires an accurate measurement of the charge of the nuclei. The 
charge $Z$ of a relativistic particle or atomic nucleus impinging on a pair-conversion telescope can be measured by exploiting the $Z^2$ 
scaling of the energy deposited by ionization on the sensitive layers. However, the presence of a non-negligible amount of passive material 
increases the fragmentation probability of the nucleus. After fragmentation, the subsequent ionization measurements are no longer useful for 
charge determination, and typically must be rejected in offline charge reconstruction by an algorithm which identifies the interaction 
point. The reduced set of independent ionization measurements and the limited accuracy of the algorithm likely lead to a worsening of 
the charge identification performance. The simplest solution to this problem is to add a dedicated sub-detector for charge measurement 
externally to the pair-conversion telescope, for example a set of silicon detectors with a minimal amount of passive material. In this way 
the charge can be precisely measured before the nucleus traverses a considerable amount of material, at the price of an increase in power 
consumption and cost.

Such a configuration consisting of a charge detector on top of a pair-conversion telescope can provide robust performance for both charge 
identification and tracking of gamma rays and charged particles, but it will have an important impact on the overall geometric factor of 
the instrument. The ability to resolve point sources with the desired accuracy requires a good tracking resolution, that translates into a 
high lever arm value for the tracking telescope (since at fixed resolution of position measurement the angular resolution scales as 
$1/L$, where $L$ is the distance between the outmost tracking layers); stacking the charge detector on top of the telescope further 
increases the overall height of the instrument. This results in a reduction of the field of view and so in a degradation of the geometric 
factor, since for a particle telescope with a distance $d$ between the entrance and exit planes defining the instrument acceptance the 
geometric factor scales approximately as $1/d^2$ \cite{Sullivan1971}. This fact will impair the possibility to detect a sufficient amount 
of high-energy particles, effectively limiting the energy range covered by the instrument.

The weight of the pair-conversion telescope can also have a detrimental impact on detector design. The mass budget of a space experiment is 
usually severely constrained, limiting the maximum weight and dimension of the calorimeter and thus the overall geometric factor. 
Allocating a significant fraction of this budget to the passive material needed for the conversion of the gamma rays further reduces the 
instrument performance.

In this paper, an alternative detector design named Tracker-In-Calorimeter (TIC) aimed at mitigating the issues described above is 
presented, together with expected performance. Section \ref{sect:measurementprinciple} describes the measurement principle; section 
\ref{sect:detectordesign} gives an example of a realistic detector design implementing the measurement principle; section 
\ref{sect:mcsimulations} shows the expected performance for a full-scale detector obtained with Monte Carlo simulations, while section 
\ref{sect:prototype} illustrates the design of a detector prototype. The results obtained using data from the prototype acquired at test 
beams are presented in section \ref{sect:tbresults}.

\section{Measurement principle}
\label{sect:measurementprinciple}

The basic idea is to use a calorimeter as a tracking device for gamma rays. It relies on the fact that most cosmic-ray detectors feature an 
electromagnetic calorimeter below the main tracking device, with the main purpose of measuring the particle energy and discriminating 
electrons and protons by means of topological shower analysis. Indeed, the effort towards the high-energy frontier in space (sub-TeV for 
gamma rays, multi-TeV for electrons and PeV for hadrons) is being pursued by running \cite{Asaoka2017,Chang2017} and future 
\cite{Zhang2017} experiments mainly with deep, homogenous electromagnetic calorimeters with scintillating crystals. When a photon hits 
the calorimeter it generates an electromagnetic shower, whose axis preserves the information about its direction. By sampling the transverse 
profile of the shower at different depths in the calorimeter it is possible to reconstruct the photon direction, e.g. by computing the 
center-of-gravity of the shower image on each layer and then fitting with a line. The accuracy and precision of this method can be improved 
by more sophisticated reconstruction algorithms and also by a more refined lateral sampling.

This tracking principle has already been implemented by other experiments, obtaining resolutions better than about 100 $\mu$m on the 
reconstructed impact point for electrons above 100 GeV using a Si-W sampling calorimeter \cite{Adriani2010}. Since the track reconstruction 
procedure relies completely on the development of an electromagnetic shower, similar performance can be expected for photons. A successful 
application of this method to the calorimetric section of a cosmic-ray detector would allow to offload the gamma-ray tracking duties to the 
calorimeter, optimising the tracker design for the tracking and possible charge measurement of primary nuclei and electrons. In practice, it 
would be possible to remove the passive material from the main tracker and also to reduce the overall height of the instrument (since 
tracking requirements for gamma rays are usually more demanding than those for charged particles due to the necessity to identify point 
sources), thus providing a solution to most of the issues described in section \ref{sect:introduction}.

\section{Detector design}
\label{sect:detectordesign}
\begin{figure}[htbp]
\begin{center}
\includegraphics[width=13cm]{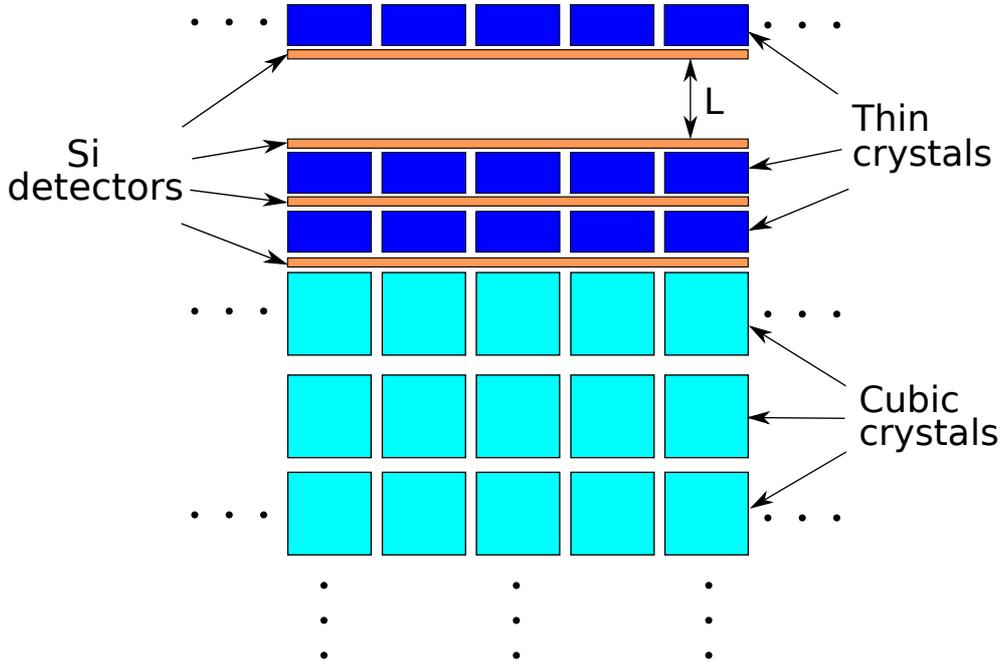}
\end{center}
\vspace{-0.5cm}
\caption{Side view of a portion of the tracking calorimeter. The first 3 scintillator layers on top are made of thin crystals (blue) 
and 
interleaved with 4 silicon microstrip layers (orange). Cubic crystals (cyan) are used below this top structure.}
\label{fig:design}
\end{figure}

This proposal for a realistic configuration of a tracking calorimeter for a future multi-messenger cosmic-ray experiment starts by 
considering the innovative proposal for a homogeneous and isotropic calorimeter made by the CaloCube collaboration 
\cite{Adriani2017,Adriani2019}. The main feature of this calorimeter is to be able to accept particles not only from the face pointing 
towards the zenith but also from lateral faces, thus boosting the geometric factor by a theoretical 5x factor in the same mass and power 
consumption envelope. This design has been chosen with some modifications for the future HERD experiment \cite{Zhang2017}, and thus 
constitutes an appealing starting point. The CaloCube sensitive elements are cubic scintillating crystal of size 1 Moli\`ere 
radius read out by two photodiodes of different sizes in order to achieve a high dynamic range. The crystals are arranged in a 
cubic 3D mesh. Such a segmented calorimeter already provides sizeable information about the development of the electromagnetic shower, 
that can be further enhanced by instrumenting one side (e.g. the zenithal one) with silicon microstrip detectors interleaved with the first 
layers of scintillating crystals, as schematically shown in figure \ref{fig:design}. Going from top to bottom, the upper layers are made of 
thin crystals rather than cubic crystals, and a gap is left between the first two silicon detectors. This arrangement reduces the crystal 
material traversed by the e$^+$-e$^-$ generated by photon conversion before hitting the precise silicon sensors, mitigating the impact of 
multiple scattering on the tracking performance. In addition to this, the small gap allows for having two consecutive position measurements 
by silicon detectors without material in between, further reducing the multiple scattering effects. The two subsequent layers of crystals 
are made of thin elements as well, providing a limited amount of material for shower development between other precise position 
measurements. Then the basic CaloCube design made of cubic scintillators is restored. The silicon detectors sample the lateral development 
of the early stage of electromagnetic showers originating in the crystals at different depths in the calorimeter with a sampling interval of 
the transverse positions much smaller than that of crystals (typically hundreds of $\mu$m against some cm). This fine-grained information is 
crucial in improving the track reconstruction performance. A similar approach has been adopted in \cite{Adriani2010}, where silicon 
detectors are interleaved with tungsten plates to build a sampling calorimeter with tracking capabilities. The novelty in this proposal is 
the use of an active material as gamma-ray converter to maximize the energy resolution. 

\section{Monte Carlo simulations}
\label{sect:mcsimulations}
To quantitatively assess the expected performance of the tracking calorimeter proposed in the previous section, a FLUKA 
\cite{Battistoni2015,Bohlen2014} Monte Carlo simulation has been set up. The simulated detector is modeled adopting some of the design 
choices of the future HERD calorimeter \cite{Zhang2017} (which in turn follows in many aspects the CaloCube design described in section 
\ref{sect:detectordesign}): the thin and cubic crystals are made of Lutetium Yttrium Orthosilicate (Lu$_{2(1-x)}$Y$_{2x}$SiO$_5$), commonly 
known as LYSO, 
and the cubes have side 3 cm while the thin scintillators are 1.5 cm (i.e. half-cube) thick. The simulated geometry is implemented after 
that depicted in figure \ref{fig:design}. The whole calorimeter is a 21x21x21 mesh of cubic crystals, with the silicon layers and thin 
crystals stacked atop the top face. The gap $L$ between the first two silicon detectors is 7 cm, while the inter-crystal gaps are 5 mm. The 
total depth along the up-down direction is $\sim$ 3.2 nuclear interaction lengths ($\lambda_I$) and $\sim$ 58 radiation lengths ($X_0$). The 
thickness and readout pitch for silicon detectors are set to 300 $\mu$m and 50 $\mu$m, respectively. The gaps between the cubes are filled 
with carbon fiber, to mimic the presence of a support structure; no other supports are present.

The simulated input spectra consist of monochromatic gamma rays at different energies (1, 10 and 100 GeV), with an isotropic angular 
distribution and a zenith angle for the initial direction limited in the $\pm$ 30$^\circ$ region around the zenith direction. The starting 
point is sampled from a flat surface just above the first layer of thin crystals, so that the whole top face is uniformly illuminated by 
incoming photons.

\subsection{Track reconstruction algorithm}
\label{sect:algorithm}
The estimator of the incident particle track is an iterative algorithm based on all the information obtained by the calorimeter signals 
(both crystals and Si detectors). The key point is that the energy of the primary photon is well known thanks to the calorimetric 
measurement, therefore all the parameters of the algorithm, which to a certain extent can depend on the energy, can be considered known 
within the energy resolution of the instrument.

The Principal Component Analysis method (PCA) has been used for the iterative track reconstruction algorithm. The first step is based on 
the crystal signals only, without using the silicon detector information. Let $N$ be the total number of crystals with signal above a 
given threshold, then the three-dimensional covariance matrix of the crystal coordinates $c_i^{(n)}$ (with $i=X,Y,Z$ and $n=1,\dots ,N$) 
weighted with the signal amplitude is:
\begin{equation}
  M_{ij}^{CAL} = \frac{1}{\sum_{n=1}^N W^{(n)}_{CAL}} \sum_{n=1}^N W^{(n)}_{CAL} ( c_i^{(n)} - C_i ) ( c_j^{(n)} - C_j ) ,
\end{equation}
where $W^{(n)}_{CAL}$ is the assigned weight of the $n$-th crystal (equal to the signal $S_{CAL}^{(n)}$ on the $n$-th crystal) and $C_i$ is 
the center of the distribution of the signals defined by:
\begin{equation}
  C_i = \frac{ \sum_{n=1}^N W^{(n)}_{CAL} c^{(n)}_i }{ \sum_{n=1}^N W^{(n)}_{CAL} }.
\end{equation}

As a first step, the eigenvector with the largest eigenvalue of the covariance matrix $M_{ij}^{CAL}$ is considered as estimator of the 
primary track. The next step is to consider the energy deposits in the silicon detectors. At a given depth $Z$, a silicon micro-strip 
detector measures only one coordinate ($X$ or $Y$), and to measure both typically two parallel detector planes arranged with perpendicular 
segmentation directions are used. For this reason, the silicon measurements along the two segmentation directions are used to independently 
reconstruct the two projections of the track on the $X-Z$ and $Y-Z$ views, and from these projections the track in the three-dimensional 
space is then obtained. In the following the procedure for $X-Z$ is described, the one for $Y-Z$ being the same.

Let $J$ be the total number of strip with signal above threshold and with $d_i^{(k)}$  (with $i=X,Z$) denote the coordinate of the $k$-th 
strip. The bidimensional covariance matrix for the PCA method is expressed by:
\begin{equation}
  M_{ij}^{SIL} = \frac{1}{\sum_{k=1}^J W^{(k)}_{SIL}} \sum_{k=1}^J W^{(k)}_{SIL} ( d_i^{(k)} - D_i ) ( d_j^{(k)} - D_j ),
\end{equation}
where $W^{(k)}_{SIL}$ is the weight for the $k$-th silicon readout channel, and with the centre of the distribution of the signals defined 
by:
\begin{equation}
  D_i = \frac{ \sum_{k=1}^J W^{(k)}_{SIL} d^{(k)}_i }{ \sum_{k=1}^J W^{(k)}_{SIL} }.
\end{equation}

To improve the angular resolution, particular care must be taken in defining the weight $W^{(k)}_{SIL}$. In the algorithm used for this 
work, 
$W^{(k)}_{SIL}$ depends on the detector plane $P$ (with $P$ ranging from 1 to the number of silicon planes) to which the strip belongs. 

Once a previous estimate of the particle track (in this case, the track obtained using only the crystal signals) is available, it can be 
affirmed that the spatial dispersion for the $X$ coordinate of its intersection in the $P$-th plane can be approximated by a Gaussian 
function. The standard deviation $\sigma_P$ can be estimated by simulation. The weighting factor is then defined by:
\begin{equation}
  W_{SIL}^{(k)} = \frac{ S^{(k)} }{ S_{P} } \times \frac{1}{\sqrt{2 \pi \sigma_P^2 }} \exp \left[ -\frac{1}{2} \left( \frac{d_i^{(k)} - 
X_P}{\sigma_P} \right)^2 \right],
\end{equation}
where $X_P$ is the intersection of the track on the $P$-th plane given by the previous estimation, $S^{(k)}$ is the signal of the $k$-th 
strip and $S_P$ is the total signal measured by the $P$-th silicon plane. The reason for the introduction of $S_P$ in the denominator is 
that the angular dispersion in the development of the electromagnetic shower increases going deeper into the calorimeter due to multiple 
scattering. Because the $S_P$ value increases as the amount of material crossed, the signals of silicon planes that lie deeper are 
weighed to a lesser extent. The track reconstruction algorithm by means of the silicon detector signals uses the PCA method with the 
covariance matrix $M_{ij}^{SIL}$. Since this new estimate depends on the previous one through the value of $W_{SIL}^{(k)}$, it can be 
repeated by using the new estimate as the previous one until convergence.

\subsection{Results}
\begin{figure}[htbp]
\begin{center}
\includegraphics[width=14.5cm]{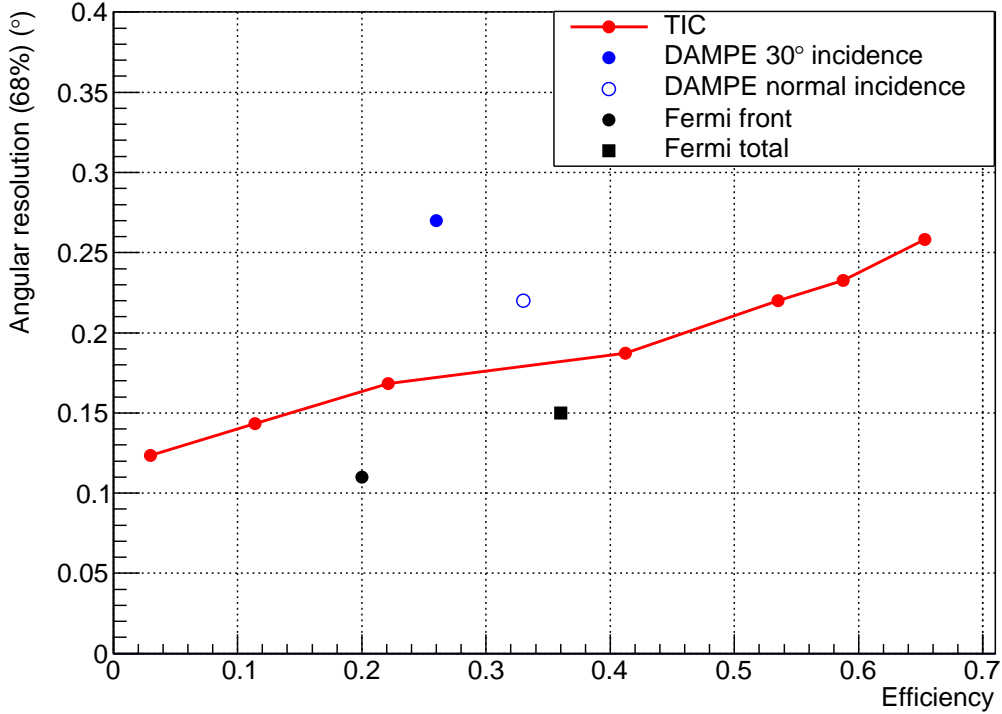}
\end{center}
\vspace{-0.5cm}
\caption{Angular resolution of TIC for 10 GeV photons as a function of the conversion+selection efficiency, compared to Fermi 
\cite{FermiPerfWebPage} and DAMPE \cite{Chang2017}.}
\label{fig:resvseff}
\end{figure}

The angular resolution has been computed for simulated photons that convert into an e$^+$-e$^-$  pair in the first thin crystal, thus 
producing a well-developed shower in the silicon detectors. The conversion efficiency in this case is $\sim$ 65\%. The resolution can be 
improved by selecting only those events for which the tracking algorithm converged within a given number of iterative steps, reducing at 
the same time the overall efficiency. This is an effective method to select ``golden samples'' of well-reconstructed events in offline 
analysis, and for tuning the accuracy according to the needed total statistics. The resolution as a function of the overall efficiency 
(conversion and selection) for 10 GeV photons is shown in figure \ref{fig:resvseff}. The TIC efficiency is the combined (conversion and 
selection) efficiency, and tops at 65\% for 100\% selection efficiency. This can be regarded as an overestimation of the real 
overall efficiency since other important factors are not considered in this estimate. For example, the trigger is expected to have a 
sizeable impact, but for implementing a realistic trigger a more refined instrument design is needed, which would reasonably include other 
sub-detectors (like an anticoincidence shield) which are unrelated to the problem of tracking gamma rays; so this aspect is neglected in 
this study. For Fermi and DAMPE the efficiency has been estimated by dividing the effective areas quoted in \cite{FermiPerfWebPage} and 
\cite{Chang2017} by the geometrical areas of the two instruments (a whole tracker plane for Fermi and the upper face of the BGO calorimeter 
for DAMPE). At this energy the resolution of TIC is worse than Fermi (total) for the same efficiency of $\sim$ 36\% but can match Fermi by 
reducing the efficiency to $\sim$ 14\%, and is generally better than DAMPE.

\begin{figure}[htbp]
\begin{center}
\includegraphics[width=14.5cm]{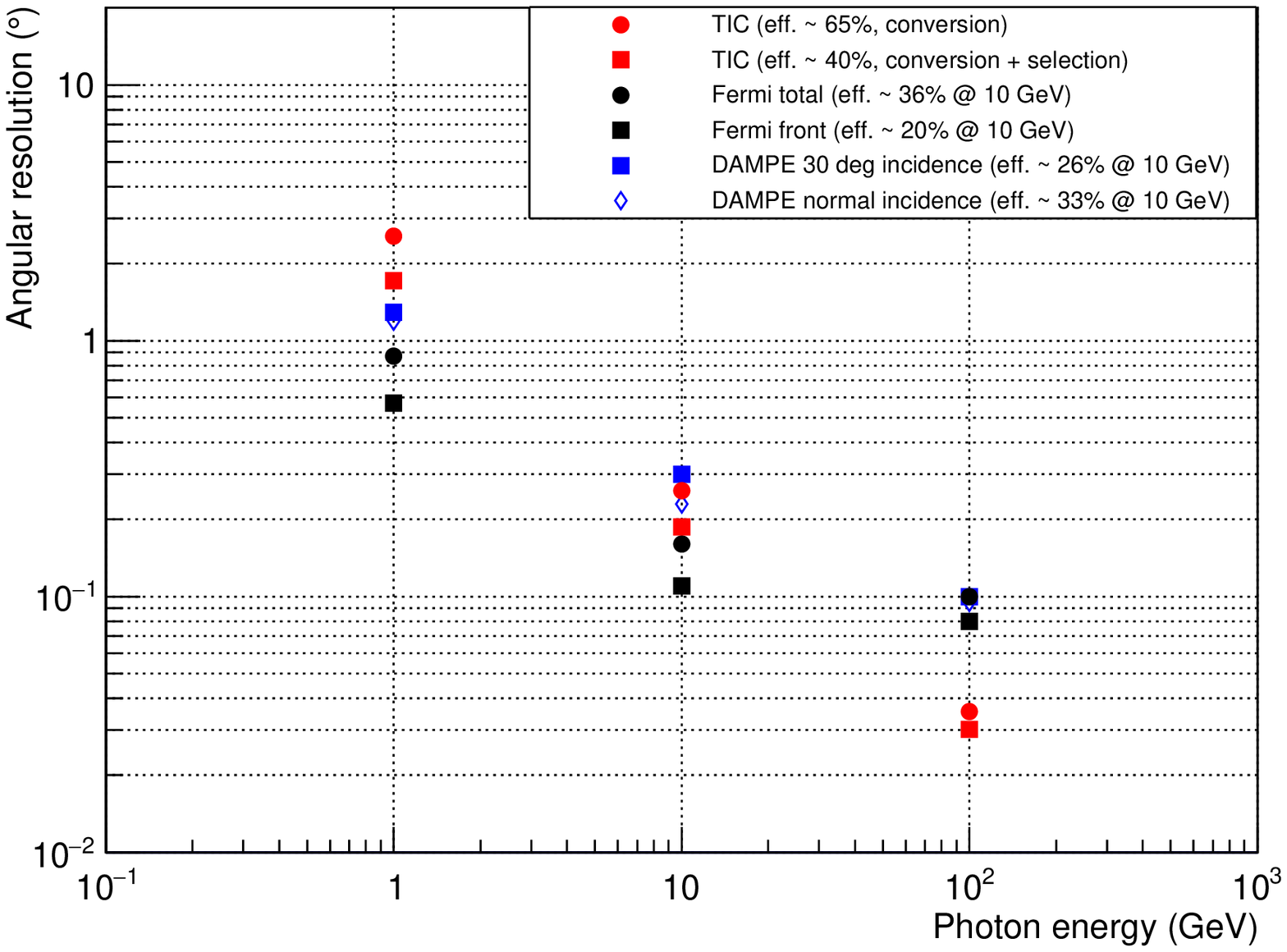}
\end{center}
\vspace{-0.5cm}
\caption{Tracking performance of TIC according to Monte Carlo simulations, compared to Fermi \cite{FermiPerfWebPage} and DAMPE 
\cite{Chang2017}.}
\label{fig:rescomparison}
\end{figure}

The performance of the tracking algorithm applied to the simulated data as a function of the gamma ray energy is compared to those of 
Fermi and DAMPE in figure \ref{fig:rescomparison}. The general trend is a performance improvement with increasing energy, which is more 
pronounced than that of Fermi and DAMPE and that provides considerably better performance above some ten GeV. In order to understand this 
behavior it is useful to consider the limiting factors for the angular resolution at different energies. In the GeV region the tracking 
precision is mainly limited by multiple scattering for both the pair conversion telescopes and TIC, the latter having thicker converters and 
thus worse performance. As the energy increases the multiple scattering becomes less relevant and the pair conversion telescopes start to 
get limited by the position resolution of their tracking systems, so that the tracking performance initially improves and then reaches a 
plateau. On the contrary, the electromagnetic showers in TIC become more populated and less affected by statistical fluctuations, so that 
the precision of the position measurements of the transverse profile continues to improve with increasing energy, and also does the tracking 
precision. At 100 GeV the performance is better than Fermi and DAMPE, with comparable or better efficiency.

\section{Detector prototype}
\label{sect:prototype}

To validate the measurement principle and the simulation results, a small-scale prototype of the tracking calorimeter was built and 
tested with particle beams. The idea behind this test was to compare the results obtained with real data with those obtained from a Monte 
Carlo simulation of the prototype detector, in order to assess the validity of the full-scale simulation results described in section 
\ref{sect:mcsimulations}. In order to reduce the construction costs, the detector was built by refurbishing the CaloCube prototype 
\cite{Adriani2019} with additional layers of thin crystals obtained by sawing some spare CaloCube cubic scintillators, and with spare 
one-sided silicon microstrip detectors from the DAMPE experiment \cite{Azzarello2016}. 

The CaloCube crystals are made of CsI(Tl), which is a different material from the LYSO used for the full-scale simulations described in 
section \ref{sect:mcsimulations}. For this reason, and also for the different sizes of the full-scale detector and the prototype, test beam 
data have to be compared to a dedicated simulation of the prototype detector and not to the full-scale simulation. The dynamic range 
required for the readout of the thin scintillators is narrower than that for cubic ones, since the former are less thick than the 
latter and are placed on the external part of the instrument where the electromagnetic showers just start to develop; hence a lower amount 
of scintillation photons with respect to the cubes is expected in thin elements. For this reason only the large-area Excelitas VTH2090 
photodiode of the high-dynamic-range, dual-photodiode readout of CaloCube has been used for thin crystals. The DAMPE silicon sensors were 
arranged in modules with two silicon wafers put side by side along the strip transversal direction; each wafer is 320 $\mu$m thick, with 384 
readout strips (for a total of 768 strips per module) and a readout pitch of 240 $\mu$m. A small dead area of $\sim$2 mm was present between 
the two active areas.

\begin{figure}[htbp]
\begin{center}
\includegraphics[width=14.5cm]{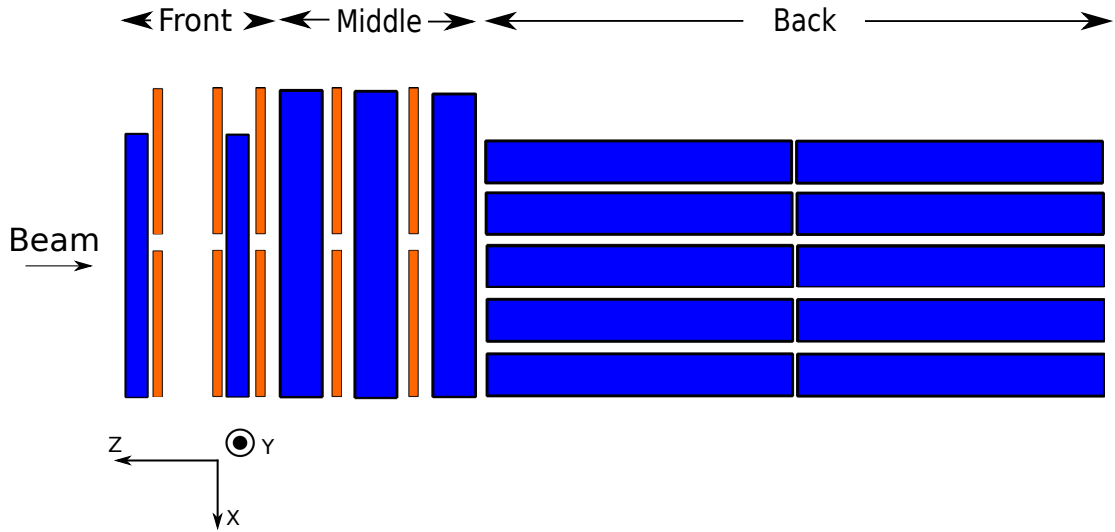}
\end{center}
\vspace{-0.5cm}
\caption{Schematic view of the prototype as seen from top. The trays containing CsI(Tl) crystals are in blue, the silicon sensors in 
orange. The beam impact angle could be changed by rotating the table on top of which the detector was mounted.}
\label{fig:prototype}
\end{figure}

The prototype was designed to resemble the scheme in figure \ref{fig:design} with some adjustments due to experimental constraints, and is 
schematically drawn in figure \ref{fig:prototype}. The {\it front} part is made of two trays of thin crystals with 5 elements along both X 
and Y directions, for a total of 25 detectors per tray with 1.7 cm thickness, and three layers of silicon microstrip detectors with one 
module on each layer. The {\it middle} part is made of three trays of 6x5 cubic elements, interleaved by two silicon layers. The front and 
middle trays are placed ortogonally to the beam. The {\it back} part is made of 10 trays of cubes identical to those of the middle part but 
arranged longitudinally with respect to the beam in two blocks of 5 trays each. All the silicon sensors have strips aligned along the Y 
direction, so that only the shower development along X can be sampled with these detectors. Indeed, due to limited availability of spare 
DAMPE sensors, it has not been possible to instrument also the Y direction. Thus, only the track projection on the XZ plane can be 
reconstructed from the silicon data, and the spatially less-precise data from the scintillating crystals must be used for the YZ projection. 
The whole structure was mounted on a rotating table which allowed to change the impact angle of the particle beam by rotating the detector 
along the Y direction.

\section{Test beam results}
\label{sect:tbresults}
The prototype was tested with electron beams of different energies at CERN test beam facilities (in May 2018 at PS and in September 
2018 at SPS). The usage of electrons instead of gamma rays is not expected to introduce sizeable differences in the estimation of the 
tracking resolution, since the tracking method is based on the sampling of electromagnetic showers which are very similar for electrons 
and photons. Furthermore, it has sizeable experimental advantages, like the possibility to have monochromatic beams of known energy up to 
$\sim$ 250 GeV and to use a very simple trigger system based on plastic scintillators read out by photomultiplier tubes.

Electron beams with energies ranging from 1 GeV to 100 GeV and incidence angles of 0 and 10 degrees were used during the two test sessions. 
The rotating table supporting the detector was set up so that a nominal rotation angle of 0$^\circ$ corresponded to an alignment of the 
detector axis with the nominal beam direction (i.e. to a nominal incidence angle of 0$^\circ$). Muons runs were used for calibrating the 
response of the instrument and for aligning all the different detector elements. 

The XZ projection of the track of the incident electrons was reconstructed with the algorithm described in section \ref{sect:algorithm}, and 
compared with the true incidence angle estimated as the mean of all the reconstructed incidence angles at fixed positioning of the rotating 
table. This procedure introduced a spread due to the different real impact angle of each electron with respect to the reference one, which 
affected the estimated angular resolution, and that was taken into account with Monte Carlo simulations as described below.

An accurate simulation of the experimental setup was realized using FLUKA, including a detailed implementation of the detector geometry and 
of several experimental effects. In the real setup the beam path was only partially in vacuum, with the electrons traveling through air 
between disconnected segments of the beam pipe, and the traversed material (air, beam pipe windows and trigger scintillators) induced the 
emission of forward-traveling bremsstrahlung photons which hit the detector and measurably affected the tracking performance. To account for 
this effect, the beam line from the first in-air segment down to the detector was implemented in the simulation. Initial particle 
kinematics was generated according to the angular and spatial distribution of the real beam as given by beam line simulations provided by 
CERN. This allowed for reproducing the spread in the angular distributions described above. Relevant instrumental effects like the 
capacitive couplings between the silicon microstrips (see \cite{Barbiellini2002} for a characterization of this effect for the silicon 
sensors used in this study) were implemented as well.

The residual rotation misalignment of the silicon sensors might induce a spread in the measured angular resolution: since the Y coordinates 
are always estimated using the scintillating crystals (due to the fact that the silicon sensors are segmented only along X) with a 
resolution of about 1 cm, this big uncertainty in determining Y can affect the determination of X when paired with the residual tilt angle 
of the silicon detectors. This effect was estimated by considering the size of the residual rotation given by the error on the 
corresponding alignment parameter as obtained from a residuals minimization procedure. When paired with a Y resolution of 1 cm this 
gives a spread of about 10 $\mu$m on X, which is small compared to the size of the residuals on X (about 50 $\mu$m at 100 GeV and 100 
$\mu$m at 10 GeV), and thus the expected impact on the angular resolution of the track reconstructed with the silicon sensors is small. This 
was verified with Monte Carlo simulations by applying a random shift on the X coordinates up to three times the expected value, with no 
sizeable effect.

\begin{figure}[htbp]
\begin{center}
\includegraphics[width=7.5cm]{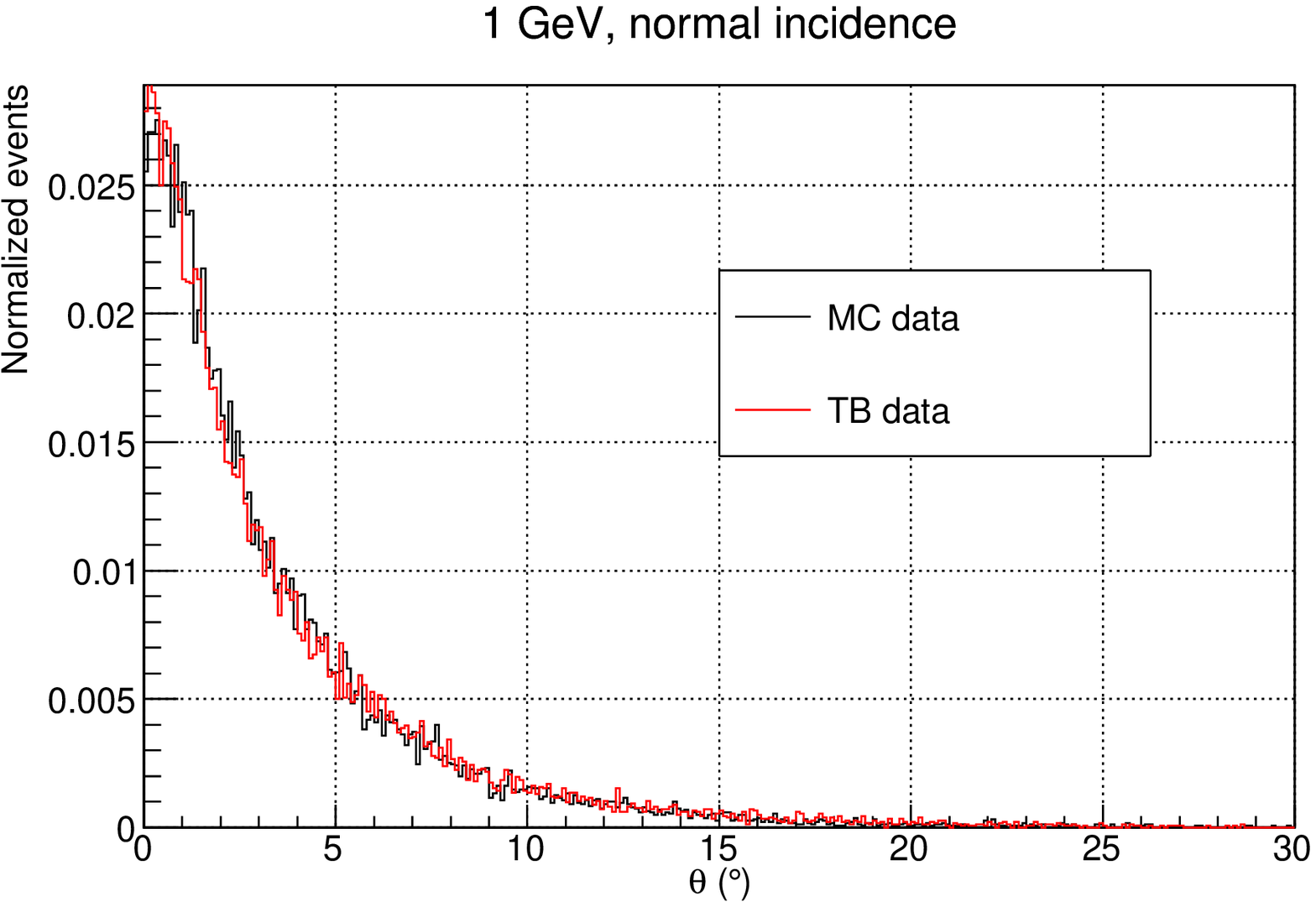}
\includegraphics[width=7.5cm]{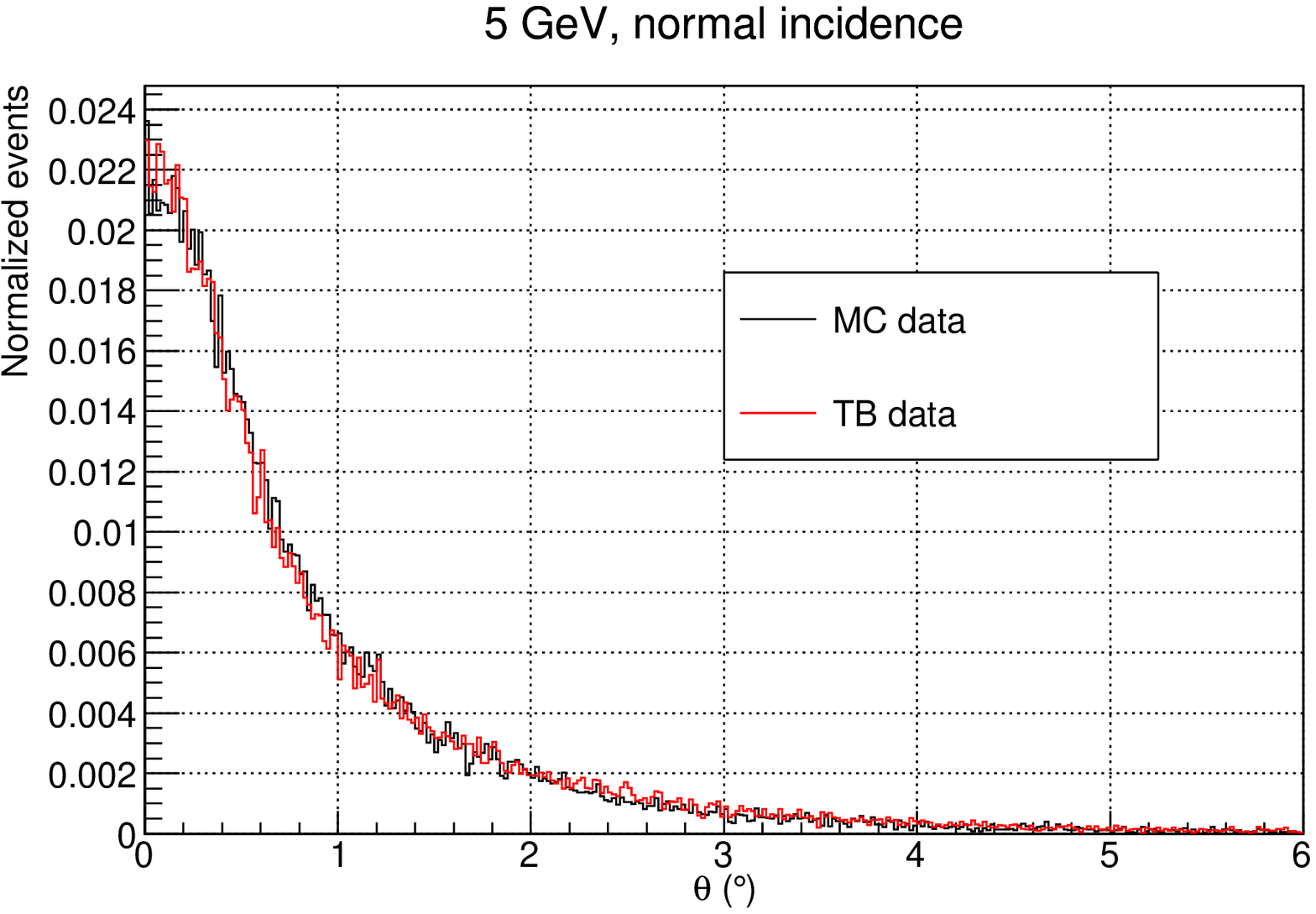}
\end{center}
\vspace{-0.5cm}
\caption{Two-dimensional tracking resolution for 1 and 5 GeV electrons with normal incidence on the detector (red: Monte Carlo simulation, 
black: test beam data).}
\label{fig:resolution1_5GeV}
\end{figure}

\begin{figure}[htbp]
\begin{center}
\includegraphics[width=7.5cm]{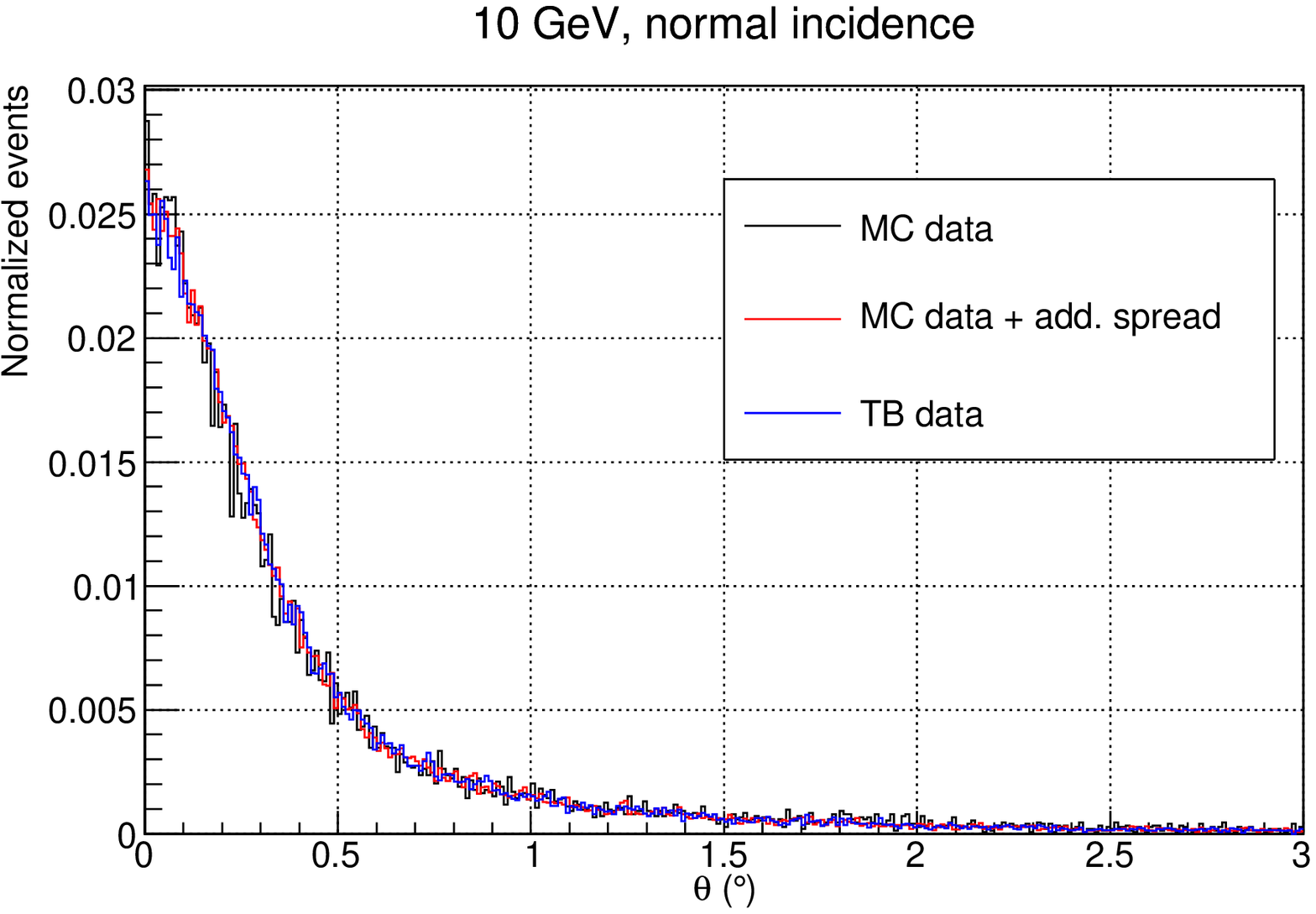}
\includegraphics[width=7.5cm]{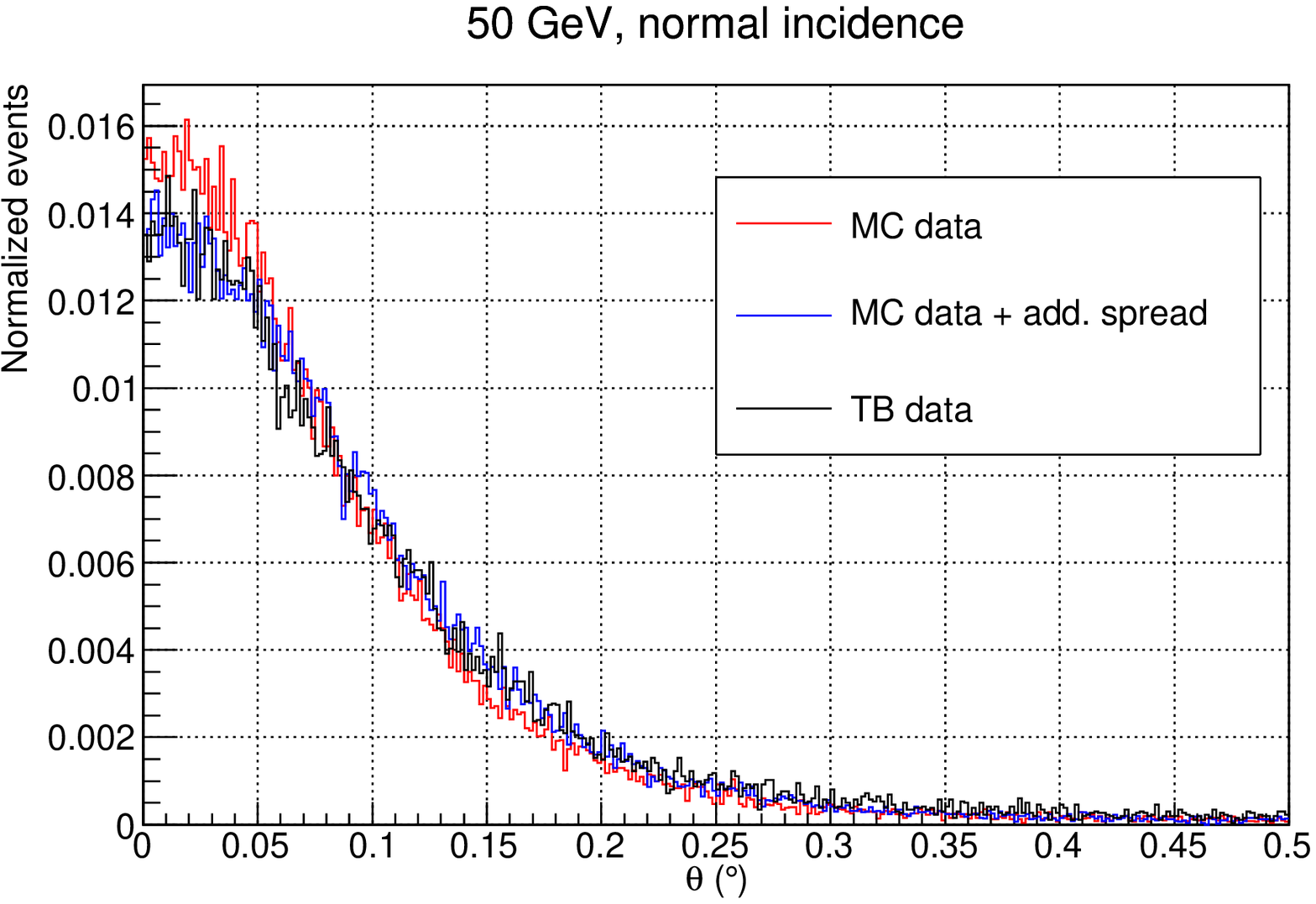}
\end{center}
\vspace{-0.5cm}
\caption{Two-dimensional tracking resolution for 10 and 50 GeV electrons with normal incidence on the detector (red: detector and upstream 
simulation with instrumental effects and beam spread, blue: same as red with additional beam spread factor, black: test beam data).}
\label{fig:resolution10_50GeV}
\end{figure}

\begin{figure}[htbp]
\begin{center}
\includegraphics[width=14.5cm]{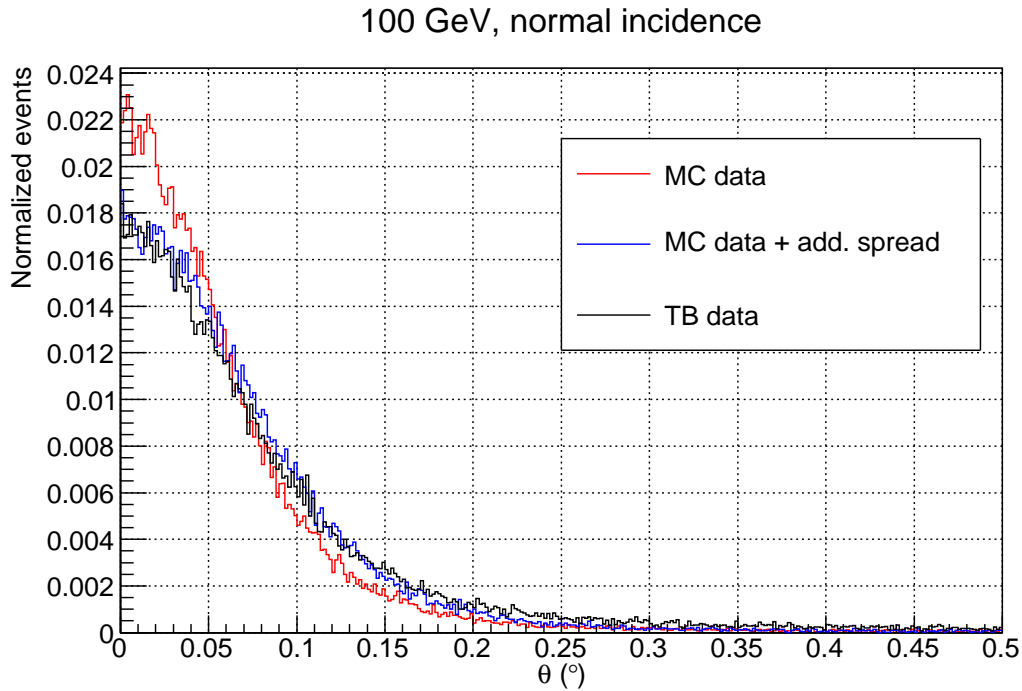}
\end{center}
\vspace{-0.5cm}
\caption{Two-dimensional tracking resolution for 100 GeV electrons with normal incidence on the detector (red: detector and upstream 
simulation with instrumental effects and beam spread, blue: same as red with additional beam spread factor, black: test beam data).}
\label{fig:resolution100GeV}
\end{figure}

The agreement with test beam data is improved by accounting for these features, and actually it is reasonable that such a careful treatment 
is needed to reproduce a very accurate angular resolution. The importance of such corrections becomes less relevant at lower energies where 
the angular resolution becomes intrinsically worse and a very good agreement between test beam and Monte Carlo data can be obtained (figure 
\ref{fig:resolution1_5GeV}). However, to fully reconcile test beam and Monte Carlo data at higher energies, an additional spread of about 
0.04$^\circ$ had to be added to the nominal spread of the simulated beam; its contribution to the overall agreement becomes more and more 
relevant as the energy increases (figures \ref{fig:resolution10_50GeV} and \ref{fig:resolution100GeV}). This correction accounts for 
unknown residual systematic effects at all energies, but since its origin was not determined then it was not used when computing the 
angular resolution from Monte Carlo data; the residual test beam - Monte Carlo resolution difference was taken as a systematic error.

\begin{figure}[htbp]
\begin{center}
\includegraphics[width=7.5cm]{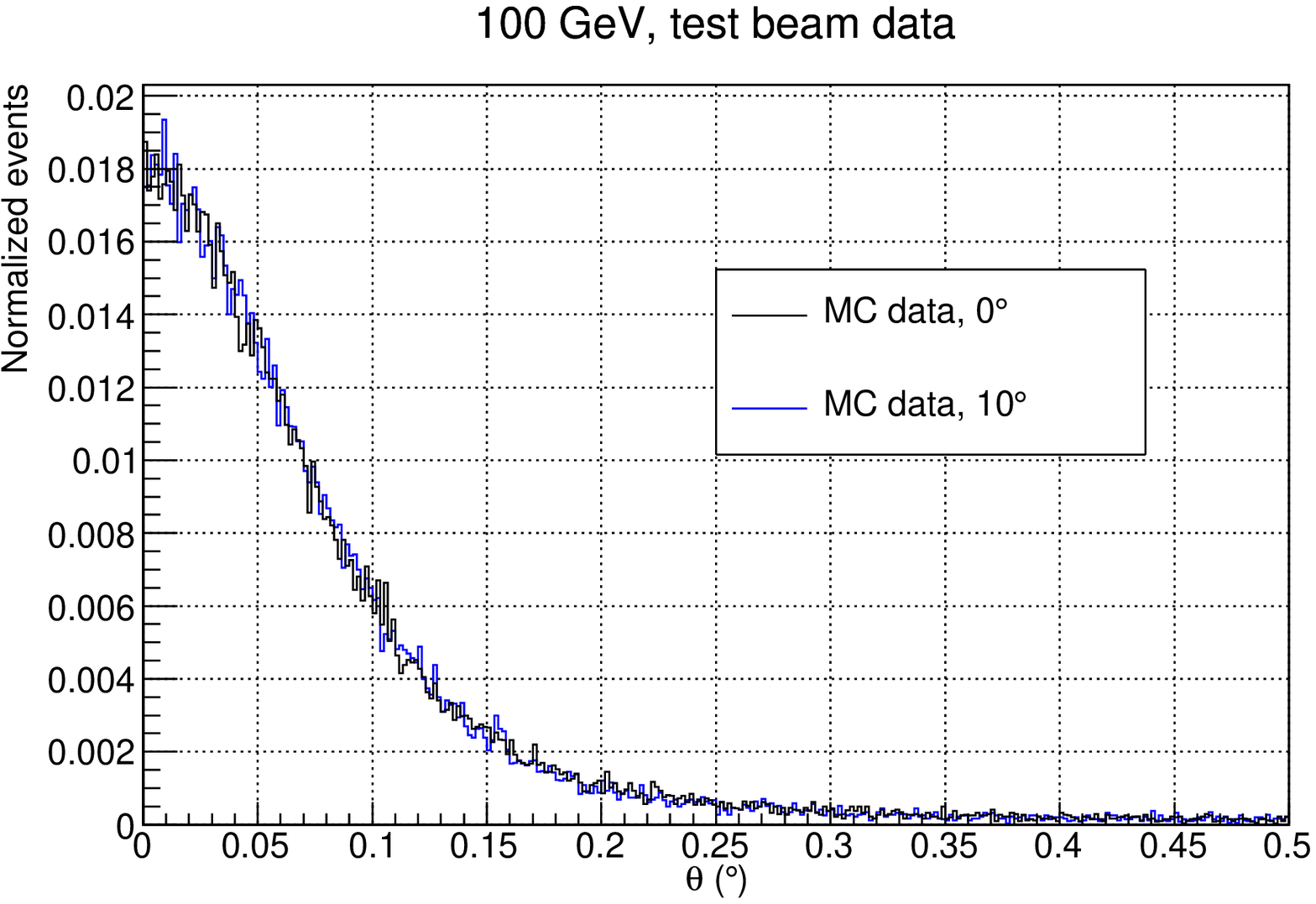}
\includegraphics[width=7.5cm]{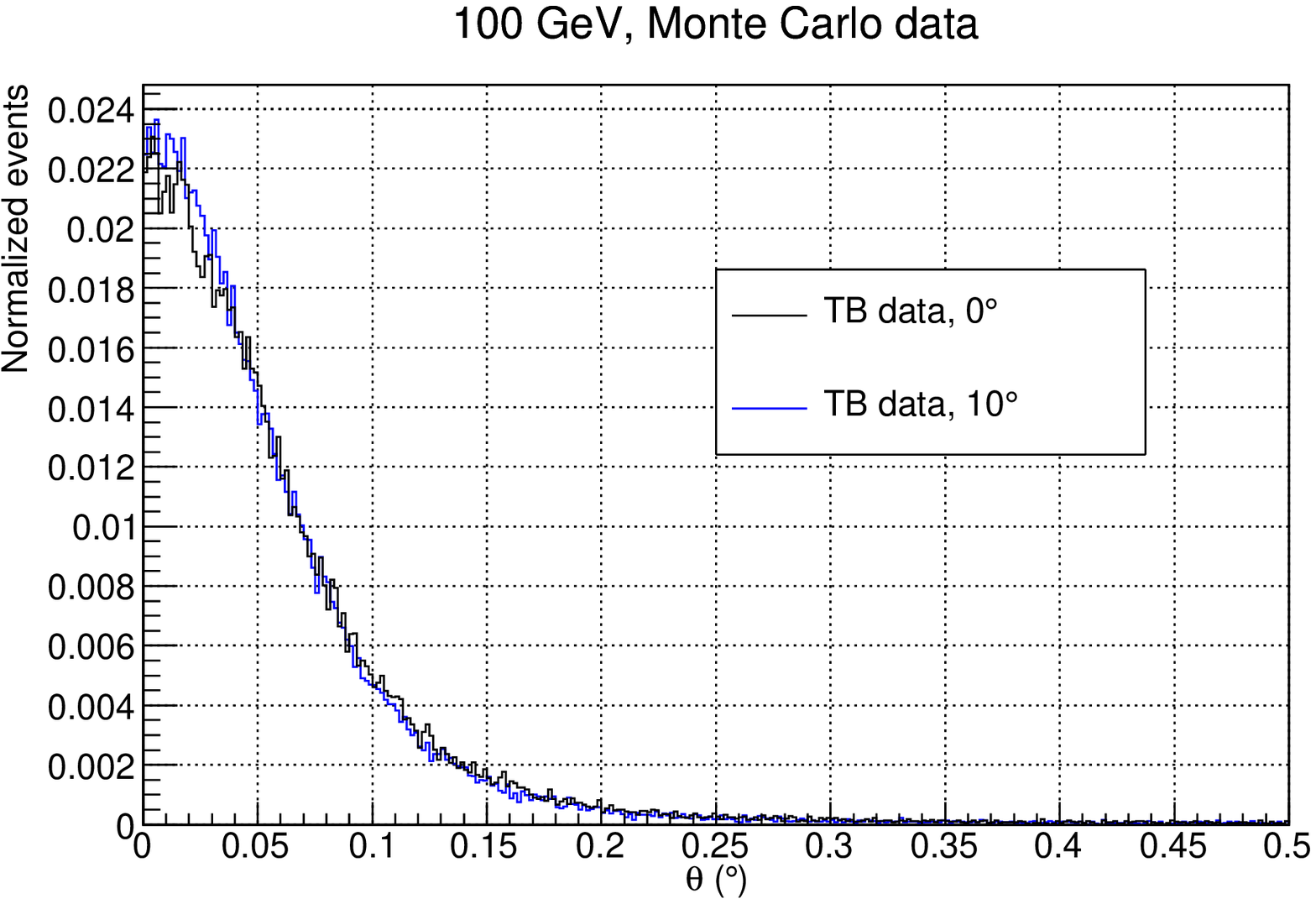}
\end{center}
\vspace{-0.5cm}
\caption{Two-dimensional tracking resolution for 100 GeV electrons for test beam (left) and Monte Carlo (right) data (black: 0$^\circ$, 
blue: 10$^\circ$ incidence angle).}
\label{fig:resolution010deg}
\end{figure}

The impact angle of the incident particle on the detector was found to have a negligible effect on the angular resolution. By comparing the 
resolution plots for 0$^\circ$ and 10$^\circ$ incidence angles no sizeable difference was found; this was confirmed by Monte Carlo 
simulations (see figure \ref{fig:resolution010deg}).

The angular resolution on the XZ projection plane are summarized in tables \ref{table:resolution68} and \ref{table:resolution95}. The 
agreement between test beam and Monte Carlo data is quite satisfactory, especially at lower energies for 68\% containment of the 
two-dimensional Point Spread Function (PSF). At 95\% PSF the agreement gets worse, likely due to residual instrumental effects that result 
in a not-so-good modeling of the tails of the distribution by the Monte Carlo. Moreover, in the test beam analysis, no particular quality 
selection criteria were applied in order reduce spurious events or the tails of the angular distribution.

\begin{table}
\begin{center}
\begin{tabular}[htbp]{|c|c|c|c|}
\hline
Energy (GeV) & Angle (deg)& Res. 68\% (deg) - TB & Res. 68\% (deg) - MC  \\
\hline
  1   &  0   & 3.72   $\pm$ 0.11    & 3.87 $\pm$ 0.12     \\
  5   &  0   & 0.985  $\pm$ 0.016   & 0.955 $\pm$ 0.014   \\
 10   &  0   & 0.4095 $\pm$ 0.0087  & 0.3971 $\pm$ 0.0050 \\
 50   &  0   & 0.1205 $\pm$ 0.0023  & 0.0995 $\pm$ 0.0050 \\
100   &  0   & 0.0897 $\pm$ 0.0010  & 0.0680 $\pm$ 0.0008 \\
100   & 10   & 0.0884 $\pm$ 0.0013  & 0.0646 $\pm$ 0.0004 \\
 \hline
\end{tabular}
\end{center}
\caption{Two-dimensional angular resolution in the XZ plane of the detector prototype, obtained from test beam (TB) and Monte 
Carlo (MC) data, for 68\% PSF.}
\label{table:resolution68}
\end{table}
\begin{table}
\begin{center}
\begin{tabular}[htbp]{|c|c|c|c|c|c|}
\hline
Energy (GeV) & Angle (deg)& Res. 95\% (deg) - TB & Res. 95\% (deg) - MC  \\
\hline
  1   &  0   & 10.63 $\pm$ 0.15  & 11.44 $\pm$ 0.33 \\
  5   &  0   & 3.300 $\pm$ 0.063 & 3.282 $\pm$ 0.054 \\
 10   &  0   & 1.946 $\pm$ 0.062 & 1.790 $\pm$ 0.068 \\
 50   &  0   & 0.921 $\pm$ 0.029 & 0.710 $\pm$ 0.058 \\
100   &  0   & 0.678 $\pm$ 0.033 & 0.3077 $\pm$ 0.0099 \\
100   & 10   & 0.662 $\pm$ 0.036 & 0.259 $\pm$ 0.019 \\
 \hline
\end{tabular}
\end{center}
\caption{Two-dimensional angular resolution in the XZ plane of the detector prototype, obtained from test beam (TB) and Monte 
Carlo (MC) data, for 95\% PSF.}
\label{table:resolution95}
\end{table}

\section{Conclusions}
In this paper, a new instrument design for tracking gamma rays in cosmic-ray space experiments has been proposed and characterized. The 
TIC tracking calorimeter has comparable tracking performance with respect to the current generation of gamma-ray experiments (Fermi, 
DAMPE) for energies from 1 GeV to $\sim$ 50 GeV, and better performance above $\sim$ 50 GeV, while allowing for important design 
optimizations for multi-particle detection. Data acquired with a detector prototype at CERN test beam facilities showed a good agreement 
with the Monte Carlo simulation, providing a deep insight on the relevant instrumental effects and a robust validation of the measurement 
principle.


\acknowledgments
This work was financed by the INFN Commissione Scientifica Nazionale 5. The authors thank F. Cadoux for helping with realizing the detector, 
and H. Wilkens, L. Gatignon, N. Charitonidis, B. Rae and M. Jeckel for test beam support.



\end{document}